\documentclass[letterpaper, 10 pt, conference]{ieeeconf}  

\IEEEoverridecommandlockouts                              
\usepackage{graphicx} 
\usepackage{epsfig} 
\usepackage{mathptmx} 
\usepackage{times} 
\usepackage{amsmath} 
\usepackage{amssymb}  
\usepackage{tikz}
\usetikzlibrary{shapes,arrows,positioning}
\usepackage{wasysym}
\usepackage{harmony}
\usepackage{musixtex}
\usepackage{pgfplots}
\pgfplotsset{compat=newest}
\pgfplotsset{plot coordinates/math parser=false}

\usepackage[colorinlistoftodos]{todonotes}
\newcommand{\noleftdelimiter}{\left.\kern-\nulldelimiterspace}
\usepackage[usestackEOL]{stackengine}
\usepackage{url}
\stackMath

\newcommand{\quotes}[1]{``#1''}

\title{\LARGE \bf
JamBot: Music Theory Aware Chord Based Generation of Polyphonic Music with LSTMs
}

\author{Gino Brunner, Yuyi Wang, Roger Wattenhofer and Jonas Wiesendanger* \\
 Department of Information Technology and Electrical Engineering\\
        ETH Z{\"u}rich\\
        Switzerland\\
        {\tt\small \{brunnegi,yuwang,wattenhofer,wjonas\}@ethz.ch}
\thanks{* 
Authors listed in alphabetical order.}
}

\begin{document}

\maketitle
\thispagestyle{empty}
\pagestyle{empty}

\begin{abstract}
We propose a novel approach for the generation of polyphonic music based on LSTMs. We  generate music in two steps. First, a chord  LSTM predicts a chord progression based on a chord embedding. A second LSTM then generates polyphonic music from the predicted chord progression. The generated music sounds pleasing and harmonic, with only few dissonant notes. It has clear long-term structure that is similar to what a musician would play during a jam session. We show that our approach is sensible from a music theory perspective by evaluating the learned chord embeddings. Surprisingly, our simple model managed to extract the circle of fifths, an important tool in music theory, from the dataset. 

\end{abstract}

\section{Introduction}

\subsection{Motivation}

Robocop, Ghost in the Shell, Titanfall: Popular culture seems to believe that robots are mechanically stronger and quicker than humans, but humans will always outsmart robots; a human mind (``ghost'') in a robot body (``shell'') is basically invincible. In the last few years, neural networks have set out to question this doctrine. While creative computing seemed out of reach not so long ago, it is getting traction with the rise of machine learning tools. Recently, neural networks have been writing novels in the style of Shakespeare \cite{KarpathyJL15}, and turning photos into paintings \cite{gatys2015neural}. 

Music is believed to be closely connected to feelings, closer than other forms of art. The area of music psychology seeks to understand the relationship between music and emotions. As such, music composition may be considered a pinnacle to understand machine creativity. 
In this work, we introduce \emph{JamBot}\footnote{Music samples and our code can be found at \url{www.youtube.com/channel/UCQbE9vfbYycK4DZpHoZKcSw} and \url{https://github.com/brunnergino/JamBot} respectively}, a music theory aware system for the generation of polyphonic music. 

Some of the early approaches to mechanically compose music using recurrent neural networks are now over two decades old \cite{bharucha1989}, \cite{mozer1994neural}. Also long short-term memory (LSTM) networks have been considered quite early \cite{lstmblues}. 
These early approaches were however limited. 
Recently models that generate polyphonic, harmonic sounding music have been proposed \cite{boulanger2012modeling}, \cite{hadjeres2016deepbach}, \cite{johnson2017generating}. 
There were also some models that integrate the concept of chords \cite{choi2016text}, \cite{songFromPi}.
In \cite{songFromPi} the monophonic melody is predicted first, and then a chord is generated and played to the melody.
Generally, these models see chords and melody as two separate entities, even though chords and melody are usually not strictly separated in music. Instead, chords and melody are two sides of the same coin: The single notes of a chord can be played like a melody, and notes of a melody can form a chord.

In contrast to other work, JamBot does not separate chords and melody. We predict the chord progression first as a structural guide for the music. Since there is only 1 chord for every 8 time steps of our polyphonic model, the chord structures last for a longer time frame; this is not possible with only one LSTM. This chord structure is then fed into a \emph{polyphonic LSTM} that generates the actual music. In contrast to other work, our polyphonic LSTM is free to predict any note, not just chord notes. The chords are only provided as information to the LSTM, not as a rule.

Our model manages to produce harmonic sounding music with a long time structure. When trained on MIDI music in major/natural minor scales with all twelve keys, our model learns a chord embedding that corresponds strikingly well to the \emph{circle of fifths}.  Thus, our LSTM is capable of extracting an important concept of music theory from the data.

\subsection{Related Work}
Neural networks have been used to generate music for decades. Mozer \cite{mozer1994neural} used a recurrent neural network that produced a pitch, duration and chord at each time step. This approach however encoded principles of music theory into the data representation. Eck and Schmidhuber \cite{lstmblues} were the first to use an LSTM. They trained the LSTM to repeat a blues chord progression, and play melodies over it.

Boulanger-Lewandowski et al. \cite{boulanger2012modeling}
proposed a model that predicts polyphonic music (multiple independent notes) with no distinction between chords and melodies, but since the predicted music is polyphonic it can form chords. The resulting music sounds pleasing and contains some long term structure. Since the music samples are a bit short it is not possible to tell if the structure spans over multiple bars.

Other approaches that create polyphonic music are Hadjeres et al. \cite{hadjeres2016deepbach}, which create nice sounding Bach chorales that always have exactly 4 voices, and Johnson \cite{johnson2017generating} which generates pleasing sounding music also with some long term structure.

Recently there have been some approaches that take chord progressions into account.
Choi et al. \cite{choi2016text} propose a text based LSTM that learns relationships within text documents that represent chord progressions.
Chu et al. \cite{songFromPi} present a hierarchical recurrent neural network where at first a monophonic melody is generated, and based on the melody chords and drums are added. It is worth noting that \cite{songFromPi} incorporates the circle of fifths as a rule for generating the chord progressions, whereas our model is able to extract the circle of fifths from the data.

Huang and Wu \cite{HuangW16} also experiment with learning embeddings for the notes. The visualized embeddings show that the model learned to distinguish between low and high pitches.

Oord et al. \cite{oord2016wavenet} created Wavenet, a text-to-speech model based on CNNs that is trained on raw audio data. They show that their model can also be used to generate music. Mehri et al. \cite{mehri2016samplernn} train hierarchical RNNs on raw audio data. Since both of these approaches use raw audio data, whereas we use MIDI files, the results are not directly comparable. Generally, systems that use MIDI files produce better sounding, less noisy music. Moreover, training on raw audio data requires more computing power, and is often infeasible with current approaches. 

It is also noteworthy that music generation models not only come from the scientific community anymore. With Avia\footnote{http://www.aiva.ai/} and Jukedeck\footnote{https://www.jukedeck.com/} two startups joined the field of neural music generation.

\section{Basics of Music Theory}
First we introduce some important principles from music theory that we use in this paper. This is a basic introduction, and we refer the reader to standard works such as \cite{mcgrain1990music} for an in-depth overview.

\subsection{Bar}
In musical notation, a \emph{bar} or \emph{measure} is a segment of time corresponding to a specific number of beats. Each beat corresponds to a note value. The boundaries between bars (hence the name) are indicated by vertical lines. In most, but not all music a bar is 4 beats long. 

\subsection{Equal Temperament}
Almost all music uses a 12 tone equal temperament system of tuning, in which the frequency interval between every pair of adjacent notes has the same ratio. Notes are:
$C,\, C\sharp/D\flat,\, D,\, D\sharp/E\flat,\, E,\, F,\, F\sharp/G\flat,\, G,\, G\sharp/A\flat,\, A,\, H$, and then again $C$ one octave higher.
One cycle (e.g., $C$ to next $C$) is called an \emph{octave}. 
Notes from different octaves are denoted with a number, 
for example $D6$ is the $D$ from the sixth octave. 

\subsection{Scale}

A \emph{scale} is a subset of (in most cases) 7 notes. 
Scales are defined by the pitch intervals between the notes of the scale. 
The most common scale is the \emph{major scale} with the following pitch intervals:
$2, \, 2,\,1,\,2,\,2,\,2,\,1$.
The first note of the scale is called the \emph{root note}. The pair of root note and scale is called a \emph{key}. The major scale with the root note $C$ contains the following notes:
$$C\xrightarrow[]{2} D\xrightarrow[]{2} E\xrightarrow[]{1} F\xrightarrow[]{2}  G\xrightarrow[]{2}  A\xrightarrow[]{2} H\xrightarrow[]{1}C.$$ 

The \emph{natural minor} scale has different pitch intervals than the major scale, but a natural minor scale with root note $A$ contains exactly the same notes as a major scale with root note $C$. We call this a \emph{relative minor}.

\subsection{Chords}
A \emph{chord} is a set of 3 or more notes played together. Chords are defined, like keys, by the pitch intervals and a starting note. The two most common types of chords are \emph{major chords} and \emph{minor chords}.  We denote the major chords with the capital starting  note, e.g., $F$ for an $F$ major chord. For minor chords we add an $m$, e.g., $Dm$ for a $D$ minor chord.

\subsection{Circle of Fifths}
\begin{figure}[t]
\centering
\includegraphics[scale=0.45]{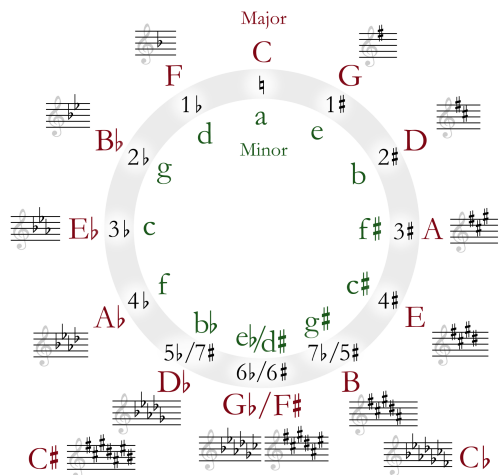}
\caption{
Circle of fifths, a visualization of the relationship between the 12 notes as it is used by musicians.}
\label{fig:circle}
\end{figure}

The \emph{circle of fifths}, which is shown in Figure \ref{fig:circle}, is the relationship among the 12 notes and their associated major and minor keys. 
It is a geometrical representation of the 12 notes that helps musicians switch between different keys and develop chord progressions. 
Choosing adjacent chords to form a chord progression often produces more harmonic sounding music.

\section{Dataset}

\subsection{General Description}

To train the models we used a subset of the Lakh MIDI Dataset \cite{lakh}. The dataset contains approximately one hundred thousand songs in the MIDI \cite{midistandard} data format. MIDI files do not contain any sounds, but rather a series of messages like \quotes{note on}, \quotes{note off}, \quotes{change tempo}. The MIDI messages are interpreted by a hard- or software MIDI instrument which then produces the sound. MIDI messages may be sent on different channels which have different sounding instruments assigned to them. For example channel 0 may represent a piano while channel 1 corresponds to a guitar.
Because MIDI files only contain a score (sheet music) of the song and no actual sound, a song usually takes much less storage space than other audio files such as WAV or MP3. 
This is also beneficial when training neural networks. 
Since the dataset is smaller, one can incorporate more songs during training. Moreover it is simple to change the instrument with which the music is played. Furthermore, the MIDI format already provides a basic representation of music, whereas a raw audio file is more difficult to interpret, for humans as well as machine learning algorithms. 

\subsection{Preprocessing}

\subsubsection{Scales and Keys}
To analyze the scales and keys of the songs we considered 5 scale types: Major, natural minor, harmonic minor, melodic minor and the blues scale. Because the major scale and its relative natural minor scale contain the same notes and only the root note is different, we treat them as the same major/relative minor scale in the preprocessing. Every scale can start at 12 different root notes, so we have $4 \cdot 12=48$ different possible keys.
To find the root notes and scale types of the songs we computed a histogram of the twelve notes over the whole song. To determine the keys, the 7 most occurring notes of the histograms were then matched to the $48$ configurations. 

Analyzing the 114,988 songs of the dataset shows that 86,711 of the songs are in the major/relative minor scale, 1,600 are in harmonic minor, 765 are in the blues scale and 654 are in melodic minor. The remaining 25,258 are in another scale, there is a key change in the song or the scale could not be detected correctly with our method. If the key changes during a song, the histogram method possibly detects neither key. Also, if a non scale note is played often in a song, the key will also not be detected correctly. 

To simplify the music generation task, we used only the songs in the major/minor scales as training data, since they make up most of the data. Additionally those songs were shifted to the same root note $C$ which corresponds to a constant shift of all the notes in a song. We call this dataset the \emph{shifted dataset} from now on. This way the models only have to learn to create music in one key instead of twelve keys. This step is taken only to avoid overfitting due to a lack of data per key. After generation, we can transpose the song into any other key by simply adding a constant shift to all the notes. If a song sounds good in one key, it will also sound good in other keys. 

Figure \ref{fig:histo} shows a histogram of all the notes in the shifted dataset. We notice that most of the notes belong to the scale, but not all of them. Therefore, simply ignoring the notes that do not belong to the scale and solely predicting in-scale notes would make the generated music ``too simplistic''. In real music, out of scale notes are played, e.g., to create tension.

\subsubsection{Range}

\begin{figure}[t]
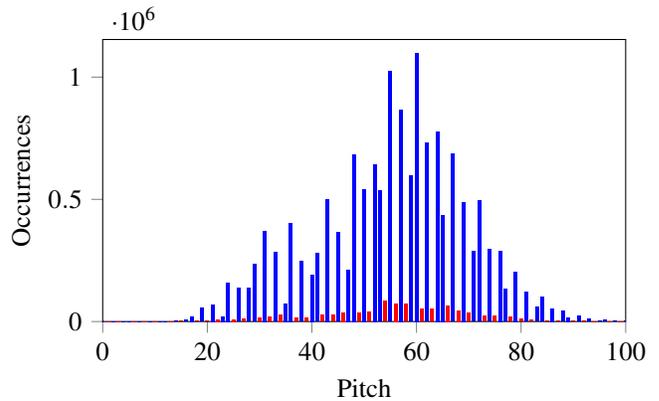

\centering
\include{figures/histo}
\caption{Histogram over the notes of the shifted dataset. The notes that belong to the $C$ major/$A$ harmonic minor scale are blue, the others red.
}
\label{fig:histo}
\end{figure}

MIDI has a capacity of 128 different pitches from $C$-1 to $G$9.
Figure \ref{fig:histo} shows that the very high and the very deep notes are not played often. Because the LSTM does not have enough data in these ranges to learn anything meaningful and the notes in these ranges usually do not sound pleasant, we only used the notes from $C2$ to $C6$ as training data.

\subsubsection{Chord Extraction}

In order to train the chord LSTM (see Section \ref{sec:chordModel}), we need to extract the chords from the songs. Because it is not feasible to determine the chords manually, we automated the process. To that end, we compute a histogram of the 12 notes over a bar. The three most played notes of the bar make up the chord. The length of one bar was chosen because usually in popular music the chords roughly change every bar.

Of course this is only an approximation to a chord as it is defined in music theory. 
We only consider chords with up to three notes, even though there are chords with four or more notes. Our method might also detect note patterns that are not chords in a music theoretical sense, but appear often in real world music. For example, if a note that is not a note of the current chord is played more often than the chord notes, the detected chord might vary from the actual chord.

In Table \ref{tab:chords} the 10 most common chords of the extracted chord datasets can be seen. In both datasets the most common chords are what one might expect from large datasets of music, and coincides with \cite{burgoyne2011expert}, \cite{chordanalysis}, \cite{spotifyinsights}. Therefore we conclude that our chord extraction method is plausible.  

\begin{table}[t]
\centering
\caption{The 10 most frequent chords in the shifted and the original dataset.}
\begin{tabular}{l|cc}
\textbf{}    & \multicolumn{1}{l}{\textbf{Shifted}} & \multicolumn{1}{l}{\textbf{Not Shifted}} \\ \hline
\textbf{1.}  & C                                    & G                                        \\
\textbf{2.}  & G                                    & C                                        \\
\textbf{3.}  & F                                    & D                                        \\
\textbf{4.}  & Am                                   & F                                        \\
\textbf{5.}  & Dm                                   & A                                  \\
\textbf{6.}  & Csus4                                & Am                                        \\
\textbf{7.}  & Em                                   & E                                       \\
\textbf{8.}  & Gsus4                                & Em                                        \\
\textbf{9.}  & Csus6                                & B                                        \\
\textbf{10.} & Asus7                                & Dm                                      
\end{tabular}
\label{tab:chords}
\end{table}

\section{Models}

\begin{figure}[t]
\centering
\scalebox{.85}{\input{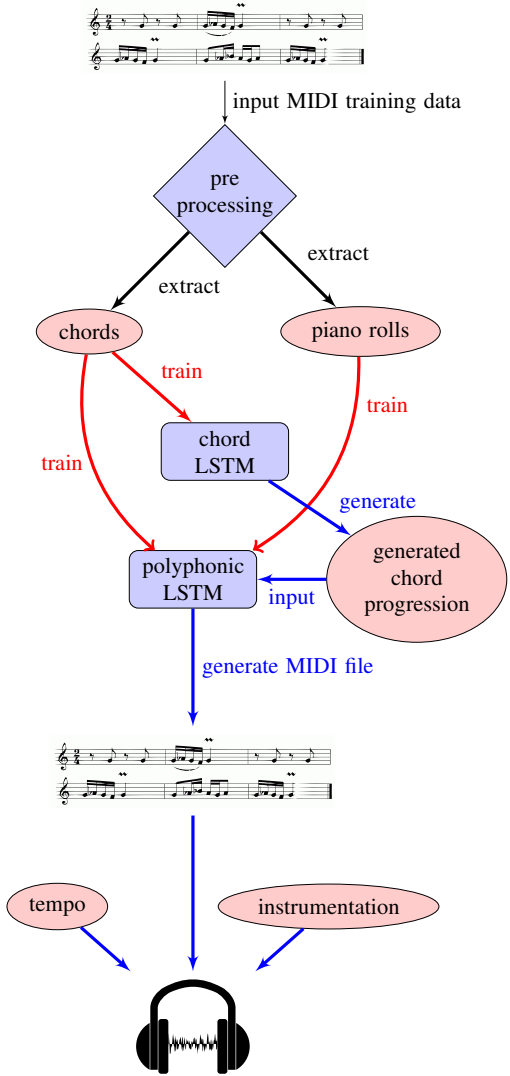}}
\caption{
The architecture of JamBot. Chords and piano roll representations are extracted from the MIDI files in the training data (in black). The extracted chords and piano rolls are then used to train the chord and polyphonic LSTMs (in red). During music generation (in blue), the chord LSTM generates a chord progression that is used as input to the polyphonic LSTM which generates new music in MIDI format. When listening to the music, one can freely vary tempo and instrumentation.}
\label{fig:architecture}
\end{figure}

When you listen to a song, dependencies in the song are important. Likewise, as you read this paper, you understand each word based on your understanding of the context and previous words. Classical neural networks, so-called \emph{Multi Layer Perceptrons} (MLP), cannot do this well. Recurrent neural networks (RNN) were proposed to address this issue, however, normal RNNs usually only capture short-term dependencies. In order to add long-term dependencies into generated music, which is believed to be a key feature of pleasing music, we use LSTM (Long Short-Term Memory) networks \cite{hochreiter1997long} which is an architecture designed to improve upon the RNN with the introduction of simple memory cells with a gating architecture. 
These gates decide whether LSTM cells should forget or persist the previous state in each loop and thus make LSTMs capable of learning useful dependencies within a long sequence. 

We denote by $x_0, \ldots, x_t,\ldots$ the input sequences and $y_0,\ldots,y_t,\ldots$ the output sequences. 
For each memory cell, the network computes the output of four gates: an update gate, input gate, forget gate and output gate.
The outputs of these gates are:
\begin{align*}
     i = \sigma(U_i x_t + V_i h_{t-1})\\
     f = \sigma(U_f x_t + V_f h_{t-1})\\
     o = \sigma(U_o x_t + V_o h_{t-1})\\
     g = \tanh(U_g x_t + V_g h_{t-1})\\
\end{align*}
where $U_i, U_f, U_o, U_g, V_i, V_f, V_o, V_g$ are all weight matrices.
The bias terms have been omitted for clarity.
The memory cell state is then updated as a function of the input and the previous state:
\[c_t = f\odot c_{t-1} + i\odot g.\]
The hidden state is computed as a function of the cell state and the output gate, and finally the output is computed as the output activation function $\delta$ of the output matrix $W_{out}$ multiplied with the hidden state:
\begin{align*}
h_t = o\odot \tanh(c_t) \\
y_t = \delta(W_{out} h_t)
\end{align*}

For more details about general LSTMs, we refer the interested readers to \cite{Goodfellow-et-al-2016}.

In Figure \ref{fig:architecture} JamBot's architecture is shown. We will explain it in detail in the remainder of this section.

\subsection{Data Representation}

\subsubsection{Polyphonic LSTM}\label{music lstm data}

To represent the music data that is fed into the polyphonic LSTM we use a \emph{piano roll} representation. Every bar is divided into eight time steps. The notes that are played at each time step are represented as a vector. The length of these vectors is the number of notes. If a note is played at that time step, the corresponding vector entry is a $1$ and if the note is not played the corresponding entry is a $0$. The piano rolls of the songs are created with the pretty\_midi library \cite{prettymidi} for Python.

\subsubsection{Chord LSTM}\label{chord lstm data}

To represent the chords of a song we borrow a technique from natural language processing. In machine learning applications that deal with language, words are often replaced with integer ids and the word/id pairs are stored in a dictionary. The vocabulary size is usually limited. Only the $N$ most occurring words of a corpus receive a unique id, because the remaining words do not occur often enough for the algorithms to learn anything meaningful from them. The rarely occurring words receive the id of an \emph{unknown tag}. For the chord LSTM we use the same technique. The chords are replaced with ids and the chord/id pairs stored in a dictionary. So the chord LSTM only sees the ids of the chords and has no knowledge of the notes that make up the chords.

\begin{figure}[t]
\centering
\begin{tikzpicture}

\begin{axis}[
xmin=-14.9, xmax=312.9,
ymin=-65490.8, ymax=1375614.8,
ylabel=Occurrences,
xlabel=Chords,
tick align=outside,
tick pos=left,
width = 0.48\textwidth,
height = 0.25\textwidth,
x grid style={lightgray!92.026143790849673!black},
y grid style={lightgray!92.026143790849673!black}
]
\addplot [semithick, blue, forget plot]
table {%
0 1310110
1 922075
2 807361
3 699285
4 351531
5 253950
6 249689
7 230997
8 203989
9 197045
10 169473
11 167128
12 156245
13 142824
14 134899
15 121724
16 119611
17 115318
18 109758
19 109295
20 103825
21 98023
22 94391
23 86729
24 80456
25 75543
26 63099
27 61559
28 58329
29 51968
30 51622
31 50388
32 46359
33 45388
34 30508
35 28053
36 27375
37 25505
38 23422
39 23363
40 22385
41 21703
42 21216
43 20420
44 20052
45 19897
46 18108
47 16806
48 16709
49 16537
50 15524
51 15447
52 15001
53 14897
54 14832
55 13721
56 11432
57 11337
58 10424
59 10370
60 10350
61 10158
62 9879
63 9632
64 9319
65 9184
66 9148
67 9106
68 9004
69 8862
70 8788
71 8669
72 8466
73 7981
74 7642
75 7596
76 7466
77 7464
78 7447
79 7296
80 7226
81 7166
82 7137
83 6653
84 6647
85 6411
86 6366
87 6341
88 6151
89 5946
90 5649
91 5602
92 5496
93 5434
94 5023
95 4785
96 4762
97 4759
98 4624
99 4602
100 4551
101 4472
102 4399
103 4362
104 4321
105 4155
106 4136
107 4116
108 4067
109 4013
110 3972
111 3970
112 3923
113 3783
114 3741
115 3716
116 3590
117 3537
118 3530
119 3483
120 3378
121 3353
122 3293
123 3149
124 3091
125 3057
126 2949
127 2949
128 2912
129 2897
130 2890
131 2852
132 2807
133 2774
134 2671
135 2636
136 2482
137 2431
138 2430
139 2410
140 2359
141 2306
142 2278
143 2268
144 2219
145 2184
146 2159
147 2132
148 2119
149 2102
150 2086
151 2041
152 2035
153 1961
154 1960
155 1954
156 1934
157 1910
158 1898
159 1876
160 1842
161 1791
162 1744
163 1743
164 1662
165 1614
166 1578
167 1502
168 1466
169 1465
170 1438
171 1413
172 1370
173 1334
174 1317
175 1290
176 1280
177 1241
178 1220
179 1217
180 1210
181 1186
182 1141
183 1108
184 1105
185 1091
186 1068
187 1022
188 1021
189 947
190 931
191 924
192 914
193 911
194 906
195 904
196 878
197 835
198 819
199 805
200 800
201 781
202 770
203 739
204 734
205 702
206 696
207 665
208 657
209 653
210 644
211 619
212 603
213 602
214 602
215 596
216 595
217 580
218 569
219 548
220 528
221 505
222 501
223 500
224 494
225 491
226 474
227 472
228 467
229 466
230 460
231 454
232 454
233 449
234 446
235 445
236 443
237 442
238 441
239 423
240 422
241 400
242 399
243 398
244 386
245 385
246 377
247 375
248 357
249 337
250 329
251 327
252 314
253 307
254 293
255 286
256 275
257 252
258 247
259 234
260 229
261 220
262 212
263 205
264 204
265 196
266 190
267 190
268 188
269 182
270 181
271 177
272 172
273 161
274 160
275 159
276 155
277 151
278 121
279 119
280 112
281 107
282 104
283 103
284 101
285 94
286 91
287 81
288 62
289 59
290 56
291 51
292 46
293 41
294 40
295 31
296 22
297 20
298 14
};
\end{axis}

\end{tikzpicture}
\caption{
This figure shows the number of occurrences of all 300 unique chords in the shifted dataset.}
\label{fig:chords}
\end{figure}
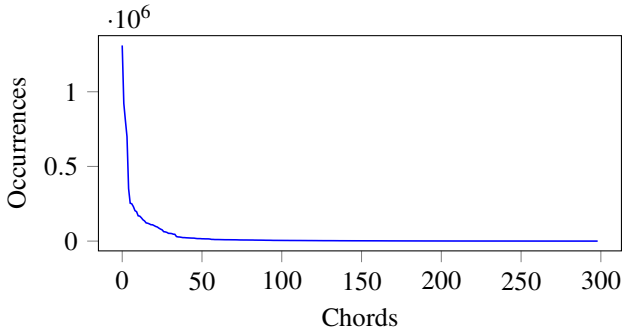

Figure \ref{fig:chords} shows the number of occurrences of all unique chords in the shifted dataset. On the left is the most frequent chord and on the right the least frequent one. Even though there are $12 \cdot 11\cdot10 + 12\cdot11 + 12 = 1,465$ different possible note combinations for 3, 2 or 1 notes, there are only 300 different combinations present in the shifted dataset. This makes sense since most random note combinations do not sound pleasing, and thus do not occur in real music. It can be seen that few chords are played very often and then the number of occurrences of the chords drops very fast. Based on this data the vocabulary size was chosen to be 50. 
The remaining chords received the id of the unknown tag. 

Before we feed the chord ids into the chord LSTM we have to encode them as vectors. To do so we use \emph{one-hot encoding}. The input vectors are the same size as the size of the chord vocabulary. All the vector entries are $0$, except for the entry at the index of the chord id which equals $1$.

\subsection{Chord LSTM}\label{sec:chordModel}
\subsubsection{Architecture}

For the first layer of the chord LSTM we used another technique from natural language processing; word embeddings. This technique has been pioneered by Bengio et al. \cite{bengio2003neural} and has since been continuously developed and improved. Google's \emph{word2vec}\cite{mikolov2013efficient} is a recent and successful result of this trend. In natural language processing, a word embedding maps words from the vocabulary to vectors of real numbers. Those embeddings are often not fixed, but learned from the training data. The idea is that the vector space can capture relationships between words, e.g., words that are semantically similar are also close together in the vector space. For example, the days of the week, or words like king and queen, might be close together in the embedding space.
For the chords we used this exact same technique. The one-hot vectors $x_{chord}$ as described in Section \ref{chord lstm data} are multiplied with an embedding matrix $W_{embed}$, resulting in a 10-dimensional embedded chord vector:
$$ x_{embed} = W_{embed} \cdot x_{chord}$$

The goal is that the chord LSTM learns a meaningful representation of the chords from the training data. In our LSTM the embedding matrix $W_{embed}$ consists of learnable parameters. Those parameters are trained at the same time as the rest of the chord LSTM.

After the embedding layer, the embedded chords are fed into an LSTM with 256 hidden cells. As output activation function softmax was used. The output of the LSTM then corresponds to a vector that contains the probabilities for all the chords to be played next.

\subsubsection{Training}
To train the chord LSTM we used cross-entropy as loss function and the Adam optimizer\cite{kingma2014adam}. The best initial learning rate we found was $10^{-5}$. The training data consists of the extracted chords of 80,000 songs from the shifted dataset. We trained the model with this data for 4 epochs. We also trained a second chord LSTM with the extracted chords of 100'000 songs from the original unshifted dataset to visualize the embeddings that it learned.

\subsubsection{Prediction}

To predict a new chord progression, we first feed a seed of variable length into the LSTM. The next chord is then predicted by sampling the output probability vector with temperature. The predicted chord is then fed into the LSTM again and the next chord is again sampled with temperature, and so on. The temperature parameter controls how divers the generated chord progression is. A temperature of zero would mean that for a given seed, the predicted chord progression would stay the same in each run.

\subsection{Polyphonic LSTM}

\subsubsection{Input}

The input vector of the polyphonic LSTM can be seen in Figure \ref{fig:input vector}. It consists of the vectors from the piano rolls of the songs, as described in Section \ref{music lstm data}, with additional features appended to the vectors.

The first feature is the embedded chord of the next time step. The embedding is the same as in the completely trained chord LSTM described in Section \ref{sec:chordModel}. With the chord of the notes to be predicted given, the LSTM can learn which notes are usually played to which chords. This way the predicted notes follow the chord progression and the generated songs receive more long term structure.

In music the melodies often \quotes{lead} to the next chord. For this reason we also append the embedded vector of the chord which follows the chord of the next time step. This way the LSTM has a target where to go with the melodies when predicting the music. This should cause the generated songs to be more structured.

The last feature that is appended is a simple binary counter that counts from 0 to 7 in every bar. This helps the LSTM to know at which time step in the bar it is and how many steps remain to the next chord change. This should make the chord-transitions smoother.

\begin{figure}
$$ 
{\bf x_{poly}^{t}} = \left( 
                                  \begin{array}{c}
                                         0\\
                                         \vdots\\
                                         3.579\\
                                         \vdots\\
                                         0.256\\
                                         \vdots\\
                                         1\\
                                         \vdots\\
                                  \end{array}
                            \right)
\setstackgap{L}{1.2\normalbaselineskip}
\vcenter{\hbox{\stackunder[1pt]{%
  \left.{\Centerstack{\\ }}\right\}\textrm{Piano roll}\hspace{2em}
}{\stackunder[1pt]{%
  \left.{\Centerstack{\\ }}\right\}\textrm{Chord}\hspace{3.5em}
}{\stackunder[1pt]{%
  \left.{\Centerstack{\\ }}\right\}\textrm{Next Chord}\hspace{1em}
}{  \left.{\Centerstack{\\ }}\right\}\textrm{Counter}\hspace{2.5em}
}}}}}
$$
\caption{The input vector of the polyphonic LSTM at time $t$. It consists of the piano roll vector, the embedded current chord, the embedded next chord and the counter.}
\label{fig:input vector}
\end{figure}

\subsubsection{Architecture}

The input vectors are fed into an LSTM with 512 cells in the hidden layer. The activation function of the output is a sigmoid. The output of the LSTM at time $t$ ${\bf y_{poly}^{t}}$ can be seen in Figure \ref{fig:output vector}. It is a vector with the same number of entries as there are notes. Every output vector entry is the probability of the corresponding note to be played at the next time step, conditioned on all the inputs of the time steps before.

\begin{figure}
$$ 
{\bf y_{poly}^{t}} = \left( 
                                  \begin{array}{c}
                                         P(n_0 = 1 | x_{poly}^0, \cdots, x_{poly}^{t-1})\\
                                         
                                         \vdots\\
                                         
                                         P(n_N = 1 | x_{poly}^0, \cdots, x_{poly}^{t-1})
                                  \end{array}
                            \right)
$$
\caption{The output vector of the polyphonic LSTM at time $t$.}
\label{fig:output vector}
\end{figure}

\subsubsection{Training}

The polyphonic LSTM is trained to reduce the cross entropy loss between the output vectors ${\bf y_{poly}^{t}}$ and the ground truth. We use the Adam optimizer with an initial learning rate of $10^{-6}$. Since for every time step in the chord LSTM there are 8 time steps in the polyphonic LSTM, the training data for the polyphonic LSTM only consists of 10,000 songs from the shifted dataset in order to reduce training time. We trained the LSTM for 4 epochs.

\subsubsection{Generation}

To predict a new song we first feed a seed consisting of the piano roll and the corresponding chords into the LSTM. The notes which are played at the next time step are then sampled from the output vector ${\bf y_{poly}^{t}}$. The notes are sampled independently, so if one note is chosen to be played, the probabilities of the other notes do not change.

We also implement a soft upper limit for the number of notes to be played at one time step. The training data mainly consists of songs where different instruments are playing at the same time with different volumes. The predicted song however is played back with only one instrument and every note is played at the same volume. So while the songs from the training data might get away with many notes playing at the same time, with our playback method it quickly sounds too cluttered. For this reason we implemented a soft upper limit for the number of notes to be played at one time step. Before prediction we take the sum of all probabilities of the output vector and if it is greater than the upper limit $l$, we divide all the probabilities by the sum and then multiply them by $l$:
$$
s = sum\{{\bf y_{poly}^{t}}\}= \sum\limits_{i=1}^N  P(n_i = 1 | x_{poly}^0, \cdots, x_{poly}^{t-1})
$$
$$
{\bf y_{poly}^{t}}_{new}= {\bf y_{poly}^{t}} \cdot (l/s)
$$
This prevents the LSTM from predicting too many notes to be played simultaneously.

In the piano roll representation there is no distinction between a note that is held for $t$ time steps and a note played repeatedly for $t$ time steps. So it is up to us how to interpret the piano roll when replaying the predicted song. We found that it generally sounds better if the notes are played continuously. To achieve this, we merge consecutive notes of the same pitch before saving the final MIDI file. However, at the beginning of each bar all notes are repeated again. This adds more structure to the music and emphasizes the chord changes.

The instrumentation and the tempo at which the predicted songs are played back with can be chosen arbitrarily. Thus, the produced music can be made more diverse by choosing different instruments, e.g., piano, guitar, organ, etc. and varying the tempo that is set in the produced MIDI file.

\section{Results}

\subsection{Chord LSTM}

\begin{figure}[t!]
\centering
\begin{tikzpicture}

\begin{axis}[
xmin=-3.5, xmax=3.5,
ymin=-3.5, ymax=3.5,
y dir=reverse,
x dir=reverse,
tick align=outside,
tick pos=left,
width = 0.42\textwidth,
height = 0.42\textwidth,
x grid style={lightgray!92.026143790849673!black},
y grid style={lightgray!92.026143790849673!black}
]
\node at (axis cs:2.42199299855028,1.30737551794545)[
  scale=0.8,
  anchor=base west,
  text=black,
  rotate=0.0
]{ D$\flat$};
\node at (axis cs:2.2347179474363,0.670889445913542)[
  scale=0.8,
  anchor=base west,
  text=black,
  rotate=0.0
]{ A$\flat$};
\node at (axis cs:2.01232272288009,-0.364774270333405)[
  scale=0.8,
  anchor=base west,
  text=black,
  rotate=0.0
]{ E$\flat$};
\node at (axis cs:1.00620924740016,-1.98901813184003)[
  scale=0.8,
  anchor=base west,
  text=black,
  rotate=0.0
]{ B};
\node at (axis cs:0.390781733021429,-2.06783862304134)[
  scale=0.8,
  anchor=base west,
  text=black,
  rotate=0.0
]{ F};
\node at (axis cs:0.153734972041467,-1.73671673794468)[
  scale=0.8,
  anchor=base west,
  text=black,
  rotate=0.0
]{ C};
\node at (axis cs:-1.5567708517112,-1.44490175458152)[
  scale=0.8,
  anchor=base west,
  text=black,
  rotate=0.0
]{ G};
\node at (axis cs:-2.30881340446967,-0.225215918133012)[
  scale=0.8,
  anchor=base west,
  text=black,
  rotate=0.0
]{ D};
\node at (axis cs:-2.1699792368138,0.876068948974796)[
  scale=0.8,
  anchor=base west,
  text=black,
  rotate=0.0
]{ A};
\node at (axis cs:-1.97538986074137,1.06544608814365)[
  scale=0.8,
  anchor=base west,
  text=black,
  rotate=0.0
]{ E};
\node at (axis cs:-1.69272248661212,1.7058989822428)[
  scale=0.8,
  anchor=base west,
  text=black,
  rotate=0.0
]{ H};
\node at (axis cs:1.48391621901843,2.20278645265375)[
  scale=0.8,
  anchor=base west,
  text=black,
  rotate=0.0
]{ G$\flat$,/F};
\end{axis}

\end{tikzpicture}
\caption{
Chord embeddings of the chord LSTM trained with the original, unshifted dataset. The learned embedding strongly resembles the Circle of Fifths. The 10 dimensional embeddings were reduced to 2 dimensions with PCA.}
\label{fig:circle plot}
\end{figure}
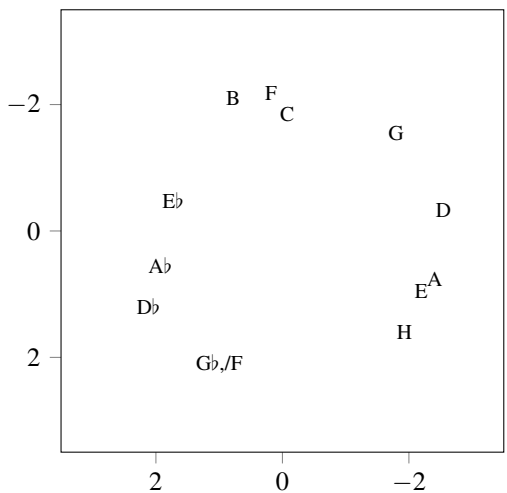

The most interesting result from the chord LSTM are the embeddings it learned from the training data. To visualize those embeddings we used PCA (Principal Component Analysis) to reduce the ten dimensional embeddings of the chords to two dimensions. 
In Figure \ref{fig:circle plot} we can see a plot of the visualized embeddings of a chord LSTM that was trained with the original unshifted dataset. The plot contains all the major chords from the circle of fifths, which we can see in Figure \ref{fig:circle}.
Interestingly the visualized embeddings form exactly the same circle as the circle of fifths. So the chord LSTM learned a representation similar to the diagram that musicians use to visualize the relationships between the chords. Thus, our model is capable of extracting concepts of music theory from songs.

In contrast to previous methods such as \cite{songFromPi} where the background knowledge is input manually to help the system do post-processing (i.e., to produce the chords with the circle of fifths), our method automatically mines this knowledge from the dataset and then exploits this mined theory to produce good songs. 
Actually, these two learning methods are also similar to the ways in which human-being learns. 
A human musician either learns the theory from her teacher, or learns by listening to a number of songs and summarizing a high level description and frequent patterns of good music. 
At a first glance, the first way appears more efficient, but in most cases encoding knowledge into a machine-readable way manually is difficult and expensive, if not impossible. 
Besides, the second learning way may help us extend the current theory by finding some new patterns from data. 

On the other hand, if someone wants to generate good music based on her own preference, but she is not an expert in music or machine learning, how could she input her own preferred \quotes{theory} into the system? Now, our data mining based method becomes more powerful since she can just tell the system which music she likes (and which not). 

This is also related to another active research field; that of learning salient representations from data. When we have a meaningful representation, similar instances should lie closely to each other in the new representation space. This phenomenon plays an important role in our model for generating high-quality new music.

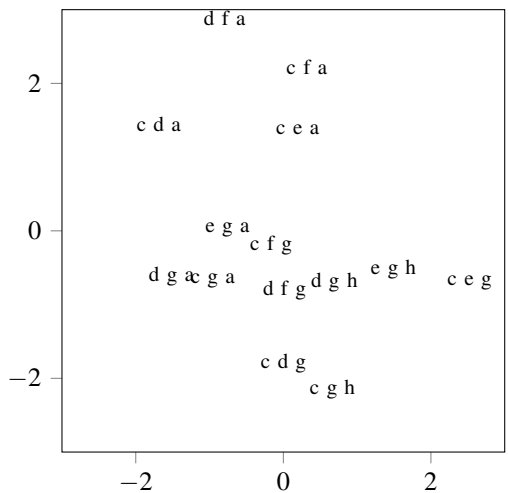
\begin{figure}[t!]
\centering
\begin{tikzpicture}

\begin{axis}[
xmin=-3, xmax=3,
ymin=-3, ymax=3,
tick align=outside,
tick pos=left,
width = 0.42\textwidth,
height = 0.42\textwidth,
x grid style={lightgray!92.026143790849673!black},
y grid style={lightgray!92.026143790849673!black}
]
\node at (axis cs:2.12503392644153,-0.730707862938615)[
  scale=0.8,
  anchor=base west,
  text=black,
  rotate=0.0
]{ c e g };
\node at (axis cs:0.275691889599173,-0.760885027728811)[
  scale=0.8,
  anchor=base west,
  text=black,
  rotate=0.0
]{ d g h };
\node at (axis cs:-0.05630627064735,2.1338177450575)[
  scale=0.8,
  anchor=base west,
  text=black,
  rotate=0.0
]{ c f a };
\node at (axis cs:-0.197478357203025,1.31476469685976)[
  scale=0.8,
  anchor=base west,
  text=black,
  rotate=0.0
]{ c e a };
\node at (axis cs:-1.17805453154424,2.79120288508573)[
  scale=0.8,
  anchor=base west,
  text=black,
  rotate=0.0
]{ d f a };
\node at (axis cs:-0.548994698579587,-0.265804518423793)[
  scale=0.8,
  anchor=base west,
  text=black,
  rotate=0.0
]{ c f g };
\node at (axis cs:1.0856934954537,-0.591736927074871)[
  scale=0.8,
  anchor=base west,
  text=black,
  rotate=0.0
]{ e g h };
\node at (axis cs:-0.40430888298249,-1.85401035202284)[
  scale=0.8,
  anchor=base west,
  text=black,
  rotate=0.0
]{ c d g };
\node at (axis cs:-1.35668469589252,-0.701239665227499)[
  scale=0.8,
  anchor=base west,
  text=black,
  rotate=0.0
]{ c g a };
\node at (axis cs:-1.15759341262144,-0.0102474276212369)[
  scale=0.8,
  anchor=base west,
  text=black,
  rotate=0.0
]{ e g a };
\node at (axis cs:-1.92220523281141,-0.676570778330125)[
  scale=0.8,
  anchor=base west,
  text=black,
  rotate=0.0
]{ d g a };
\node at (axis cs:0.258969030716281,-2.21126342888301)[
  scale=0.8,
  anchor=base west,
  text=black,
  rotate=0.0
]{ c g h };
\node at (axis cs:-0.376895672455349,-0.870799891850795)[
  scale=0.8,
  anchor=base west,
  text=black,
  rotate=0.0
]{ d f g };
\node at (axis cs:5.54020838286215,1.07994188450664)[
  scale=0.8,
  anchor=base west,
  text=black,
  rotate=0.0
]{ };
\node at (axis cs:-2.08707497033543,1.35353866859197)[
  scale=0.8,
  anchor=base west,
  text=black,
  rotate=0.0
]{ c d a };
\end{axis}

\end{tikzpicture}
\caption{
Chord embeddings of the chord LSTM trained with the shifted dataset. Instead of the chord names, the notes of the chord are shown. The ten dimensional embeddings were again reduced to two dimensions with PCA.}
\label{fig:chord notes}
\end{figure}

In Figure \ref{fig:chord notes} we used the same technique to visualize the chord embeddings trained on the shifted dataset. The embeddings of the 15 most occurring chords are plotted. Instead of the chord names the three notes that make up each chord are shown. We can see that chords which contain two common notes are close together. It makes sense that chords that share notes are also close together in the vector space. The circle of fifths is not present in the chord LSTM trained with the shifted dataset. Not even all chords are present in the chord dictionary, since its size has been limited to 50. This makes sense because many of the those chords do not occur often in $C$ major/$A$ harmonic minor.

The chord progressions predicted by the chord LSTM contain structures that are often present in western pop music. It often repeats four chords, especially if the temperature is set low. If the temperature is set higher, the chord progressions become more divers and there are fewer repeating structures.
If the sampling temperature is low, the predicted chords are mostly also the ones that occur the most in the training data, i.e., from the Top 10 in Table \ref{tab:chords}. If the sampling temperature is high the less occurring chords are predicted more often.

\subsection{Polyphonic LSTM}

The songs generated by the polyphonic LSTM sound pleasing. There clearly is a long term structure in the songs and one can hear distinct chord changes. The LSTMs succeeded in learning the relationship between the chords and which notes can be played to them. Therefore it is able to generate polyphonic music to the long term structure given by the predicted chords.

The music mostly sounds harmonic. Sometimes there are short sections that sound dissonant. That may be because even if the probabilities for playing dissonant notes are small, it can still happen that one is sampled from time to time. Sometimes it adds suspense to the music, but sometimes it just sounds wrong.

With a lower sampling temperature for the chord LSTM, the songs sound more harmonic but also more boring. Accordingly, if the sampling temperature is high, the music sounds less harmonic, but also more diverse. This might be because the chord LSTM predicts more less occurring chords with a higher temperature and there is less training data to learn the relationship between the less occurring chords and the notes.

\section{Conclusion and Future Work}
\subsection{Conclusion}
We introduced JamBot, a system to predict chord progressions as a structural guide for a song and then generate polyphonic music to those chord progressions. The generated music has a long term structure similar to what a human musician might play during an improvisation (\quotes{jam}) session. 

By visualizing the embedded chords, we show that JamBot learns the circle of fifths from the original dataset. When trained with the shifted dataset it also learns meaningful embeddings, where related chords are closer together in the embedding space. This is especially surprising considering that the chord LSTM only was provided with the chord ids. It did not receive any information about the notes of the chords. Thus, without having to explicitly implement principles of music theory, the model gained an understanding of them by observing them in the dataset.

JamBot has a simple structure and is thus easy to implement and use. Since it uses MIDI data instead of raw audio files it is fast to train on a single GPU. 

\subsection{Future Work}

JamBot is capable of learning meaningful representations. We plan to incorporate more representation learning methods, such as autoencoders, in order to learn more complex music theory related representations from the data.

The notes to be played in the next time step are sampled independently from the probability output vector ${\bf y_{poly}^{t}}$ (Figure \ref{fig:output vector}). However, it matters which notes are played together, 
since the intervals between the played notes characterize the chords and harmonies. 
This is a common problem for models that generate polyphonic music. 
We were able to mitigate this problem by providing the polyphonic LSTM with the current chord. 
Instead of sampling every note probability $P(n_i=1)$ independently, one could come up with a way to calculate the joint probabilities $P(n_0, \cdots, n_N)$ of the notes. 
This could help reduce the number of dissonant notes and would be closer to how humans compose music. 

One limitation of the piano roll data representation is that it cannot distinguish between a note that is held for several time steps and a note that is repeatedly played at every time step. Most existing data representations that address this problem only work for monophonic music. A data representation that allows both polyphony and notes of different lengths would be favorable.

So far we use a two-level approach: In the first step we generate chord progressions. In the second step the generated chords are used to generate music. Thus, the chord LSTM guides the polyphonic LSTM and helps it produce music with long-term structure. It would be interesting to add more levels to the hierarchy, by for example adding another network that guides the chord LSTM. This might enable our system to produce music with repeating structures such as choruses and verses. 

Parts of the Lakh MIDI dataset are aligned with the Million Song Dataset \footnote{\url{https://labrosa.ee.columbia.edu/millionsong/}} that contains meta information like artist, genre and lyrics of the songs. To make the generated music more diverse, one could input a \quotes{genre feature} into the LSTMs. When generating a new song one could provide the LSTMs with any desired genre feature, thus conditioning its output on said genre.

\bibliography{references.bib}
\bibliographystyle{IEEEtran}

\addtolength{\textheight}{-12cm}   






\end{document}